\documentclass[twocolumn,american,floatfix, table, dvipsnames,superscriptaddress]{revtex4-1}
\usepackage[T1]{fontenc}
\usepackage[latin9]{inputenc}
\usepackage{color}
\usepackage{babel}
\usepackage{amsmath}
\usepackage{amssymb}
\usepackage{graphicx}
\usepackage{esint}
\usepackage[unicode=trueam,
 bookmarks=false,
 breaklinks=true,pdfborder={0 0 1},backref=false,colorlinks=true]
 {hyperref}
\hypersetup{
 pdfcreator={},pdfproducer={LaTeX with hyperref},linkcolor=blue,anchorcolor=blue,citecolor=blue,filecolor=red,menucolor=red,pagecolor=red,urlcolor=blue,pdfstartview=FitV,pdfhighlight=/I,pdfpagelayout=TwoColumns,hypertexnames=true}

\makeatletter
\usepackage{lmodern}
\usepackage{mathptmx, newtxtext, newtxmath, xspace}
\usepackage{bbm}

\usepackage[table, dvipsnames]{xcolor}
\definecolor{colorA}{cmyk}{0,0,0,0.05}
\definecolor{colorB}{cmyk}{0.14,0.04,0,0}
\definecolor{colorC}{cmyk}{0.02,0.0799,0,0}
\definecolor{colorD}{cmyk}{0.099,0.14,0,0}

\usepackage{natbib}
\setcitestyle{numbers}

\usepackage{booktabs}
\usepackage{dashrule} 

\makeatother

\begin{document}
\title{Superconductivity of Incoherent Electrons near the Relativistic Mott Transition \\ in Twisted Dirac Materials}
\author{Veronika C. Stangier }
\affiliation{Institute for Theory of Condensed Matter, Karlsruhe Institute of Technology,
Karlsruhe 76131, Germany}
\author{Mathias S. Scheurer}
\affiliation{Institute for Theoretical Physics III, University of Stuttgart,  Stuttgart 70550, Germany}
\author{Daniel E. Sheehy }
\affiliation{Department of Physics and Astronomy, Louisiana State University, the
Baton Rouge, LA 70803 USA}
\author{J{\"o}rg Schmalian }
\affiliation{Institute for Theory of Condensed Matter, Karlsruhe Institute of Technology,
Karlsruhe 76131, Germany}
\affiliation{Institute for Quantum Materials and Technologies, Karlsruhe Institute
of Technology, Karlsruhe 76131, Germany}
\date{\today }
\begin{abstract}
We demonstrate that superconductivity driven by strong quantum-critical fluctuations can emerge near relativistic Mott transitions in twisted two-dimensional materials, taking on a remarkably rich character. In twisted double-bilayer WSe$_2$, all time-reversal-even, gap-opening collective modes promote pairing, whereas time-reversal-odd modes do not. In a Dirac model of twisted bilayer graphene, the Gross-Neveu transition into inter-valley-coherent insulators gives rise to a spectrum of degenerate and nearly degenerate superconducting states. More generally, we show that the richer the Dirac structure, the more readily pairs can form.  A crucial ingredient of the theory is that critical fluctuations render the electronic states strongly incoherent, allowing attractive pairing channels to overcome the bare Dirac semi-metal behavior. Finally, we demonstrate a direct relation between boson-mediated pairing and the formation of charge-carrying skyrmionic excitations in the proximate insulating state.
\end{abstract}
\maketitle
Twisted two-dimensional materials have emerged as a versatile setting
to study the tunable interplay of electron correlations, unconventional order and topology~\cite{li2010,cao2018,cao2018b,andrei2020,Yankowitz2019,Lu2019,Xie2019,Kerelsky2019,Serlin2020,Zondiner2020,Balents2020,Kennes2021,Ma2024}. Through the control of twist angle, carrier density,
and displacement fields, these systems can be driven into a remarkable
variety of ordered and topological states, ranging from correlated
insulators to superconductors. 
At the single-particle level many twisted materials possess symmetry-protected massless Dirac points. Prominent examples, where the Dirac point is right at the Fermi level,  include twisted bilayer graphene (TBG) at charge neutrality~\cite{Kerelsky2019,Zondiner2020} or stacked twisted double-bilayer WSe$_{2}$ at filling $\nu=2$~\cite{Angeli2021,Pan2023,Foutty2023,Ma2024,Tolosa2025}. 
Increasing correlations, e.g. by changing the twist angle, tends to increase correlation effects and even open a gap at the Dirac points~\cite{Lu2019,Xie2019,Choi2019,Ma2024,Stepanov2020,Xiao2025}. 
The description of this semimetal-insulator transition in terms of the Gross-Neveu (GN) mechanism~\citep{Gross1974,ZinnJustin1991} was shown to yield quantitative agreement for e.g. the value of the twist angle of the transition~\cite{,Parthenios2023,Biedermann2025,Hawashin2025,Huang2025}. 
For such a  relativistic Mott transition, the critical
point represents the onset of spontaneous symmetry breaking that gaps
the Dirac spectrum and is governed by strong, quantum-critical fluctuations.

\begin{figure}
    \centering
    \includegraphics[width=0.95\linewidth]{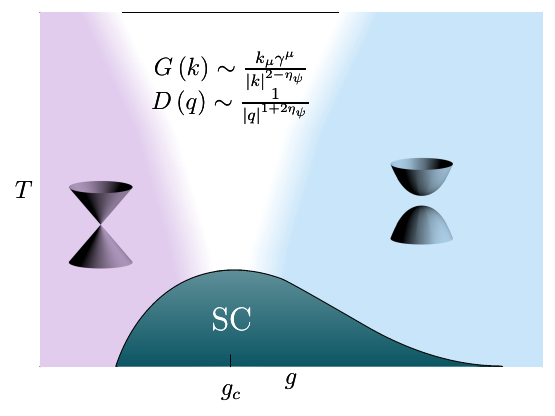}
    \caption{Schematic phase diagram for the behavior near the Gross-Neveu mass-generating transition (at $g_c$) of two-dimensional Dirac systems within the approach of Ref.~\cite{Stangier2025}. For systems with a sufficiently large anomalous fermion dimension, $\eta_\psi$, in the electron Green's function $G(k)$, critical
    boson fluctuations (with propagator $D(q)$)
    give rise to  superconductivity (SC) even in the absence of a
    sharp Fermi surface or any well-defined charge carriers.}
    \label{fig:phaseDiagram}
\end{figure}

In a parallel manuscript~\cite{Stangier2025}, we analyzed the conditions under which quantum-critical fluctuations near GN criticality can induce superconductivity at the neutrality point of a generic two-dimensional Dirac system. As illustrated in Fig.~\ref{fig:phaseDiagram}, strong-coupling superconductivity can emerge even though only a small number of carriers are thermally or quantum-mechanically excited. The nature of the pairing state is highly sensitive to the underlying critical bosonic mode. Remarkably, superconductivity occurs only when the fermions are sufficiently incoherent: well-defined quasi-particles remain non-superconducting, whereas strongly renormalized, ill-defined fermions are able to condense. An important open question is how the relativistic Mott transition in concrete twisted systems intertwines with this mechanism of superconductivity.

In this work, we investigate superconductivity in the vicinity of relativistic Mott transitions in twisted two-dimensional materials. To this end, we generalize and apply the theoretical framework of Ref.~\cite{Stangier2025} to the concrete cases of twisted double-bilayer WSe$_{2}$ and twisted bilayer graphene (TBG). For twisted WSe$_{2}$, we demonstrate that all time-reversal-even fluctuations promote pairing, whereas time-reversal-odd fluctuations do not induce superconductivity. In TBG, focusing on inter-valley coherent insulators~\cite{Po2018,Bultinck2020,Liu2021}, we uncover a rich spectrum of degenerate and nearly degenerate superconducting states. Thus, precisely at the twist angle where the fermions become gapped, one should expect either superconductivity or, at the very least, strong pairing fluctuations. Furthermore, we establish a direct connection between our framework and the intriguing possibility of charged skyrmions as a topological mechanism for pairing~\cite{Wiegmann1999,Abanov2000theta,Grover2008}, as has been intensely discussed in the context of TBG~\cite{Christos2020,Khalaf2021charged,Ledwith2021strong}.


We are interested in the solution of the two-dimensional
coupled Dirac-fermion boson problem with Hamiltonian 
\begin{eqnarray}
H & = & v_{{\rm F}}\int d^{2}x\psi^{\dagger}\left(\boldsymbol{x}\right)\left(-i\nabla\right)\cdot\boldsymbol{\alpha}\,\,\psi\left(\boldsymbol{x}\right)
\nonumber \\
 & + & \frac{1}{2}\int d^{2}x\sum_{a}\left(\pi_{a}^{2}\left(\boldsymbol{x}\right)+\omega_{0}^{2}\phi_{a}^{2}\left(\boldsymbol{x}\right)+v_{{\rm B}}^{2}\left(\nabla\phi_{a}\left(\boldsymbol{x}\right)\right)^{2}\right)\nonumber \\
 & + & g\sum_{a}\int d^{2}x\left(\psi^{\dagger}\left(\boldsymbol{x}\right)\Upsilon_{a}\psi\left(\boldsymbol{x}\right)\phi_{a}\left(\boldsymbol{x}\right)+h.c.\right).\label{eq:Hamiltonian}
\end{eqnarray}
Here $\psi$ is an $n_{\gamma}$-component Dirac spinor with  $\left\{ \alpha_{i},\alpha_{j}\right\} =2\delta_{ij}$ and $\phi_{a}$
is an $m_{\phi}$-component critical bosonic field ($a=1\cdots m_{\phi}$). $\pi_a$ is its conjugate momentum. $\omega_{0}$ is the bare mass of the boson, $v_{{\rm F},{\rm B}}$ are the fermion and boson velocities, respectively,  and $g$ is the coupling constant, where the symmetry of the boson determines the $n_\gamma \times n_\gamma$ matrices $\Upsilon_a$ that act in spinor space and that have a common parity $\tau_\phi=\pm$ under time reversal.

To address the  strong coupling behavior of this model at
the GN critical point, we  utilize a generalized  Sachdev-Ye-Kitaev
approach~\cite{sachdev1993,georges2000,sachdev2010,kitaev2015}. Following Refs.~\cite{Esterlis2019,wang2020,Kim2021,Stangier2025} we
introduce additional $N$ fermion and $M$ boson flavors,  randomize
the coupling constant $g$ between these additional flavors, and  
perform a controlled large $N$ and $M$ expansion at fixed $M/N$. This theory was formulated for the normal state  in Ref.~\cite{Kim2021}.
 The crucial aspect of the approach  is that so-called melonic diagrams dominate the physical behavior. In the normal state  and at the GN-critical point one obtains universal power-law behavior for the critical fermions and bosons (see Fig.~\ref{fig:phaseDiagram}), 
 with a comparatively large anomalous fermion dimension $0<\eta_{\psi}<\tfrac{1}{2}$ within a controlled theory. 

 To determine possible superconducting states we analyze the anomalous self energy $\Phi(k)$ in the Nambu-Gor'kov description of superconductors, which is also an $n_\gamma\times n_\gamma$ matrix in spinor space~\cite{Stangier2025}. The linearized equation for $\Phi(k)$ is:
 \begin{equation}
\Phi\left(k\right)=\frac{\lambda_{p}\tau_\phi}{m_{\phi}}\sum_{a=1}^{m_{\phi}}\int\frac{d^{3}p}{4\pi}\frac{p_{\mu}p_{\nu}\Upsilon_{a}\gamma^{0}\gamma^{\mu}\Phi\left(p\right)\gamma^{\nu}\gamma^{0}\Upsilon_{a}}{\left|p\right|^{4-2\eta_{\psi}}\left|k-p\right|^{1+2\eta_{\psi}}}.
\label{Eq:linearizedEquation}
\end{equation}
 At the GN-critical point, the power-law structure of the kernel and the dimensionless pairing strength
\begin{equation}
\lambda_{p}=\frac{2^{1+2\eta_{\psi}}\left(3-\eta_{\psi}\right)\Gamma\left(2-\eta_\psi\right)\Gamma\left(\tfrac{1}{2}+\eta_\psi\right)\sin\left(\pi\eta_{\psi}\right)}{2\pi^{3/2}\left(1+\eta_{\psi}\right)}
\label{eq:lambda_p}
\end{equation}
are universally determined by the anomalous exponent $\eta_{\psi}$. The power laws are due to the critical behavior of the fermions and bosons while $\lambda_{p}$ determines the degree to which the theory is strongly coupled.
It vanishes if $\eta_{\psi}\rightarrow0$ and is of order unity if
$\eta_{\psi}$ is no-longer small (it reaches $\frac{5}{3\pi}$ for
$\eta_{\psi}\rightarrow\tfrac{1}{2}$). $\lambda_{p}$ plays a role similar
to the  't Hooft coupling~\citep{Hooft1974} in the study
of strongly coupled gauge theories, especially in the large-$N$ limit. 

The anomalous self energy determines the usual pairing function $\Delta(k)_{ab}\sim \langle \psi_a \psi_b\rangle$ via $ \Delta\left(k\right)=\Phi\left(k\right)u_T$, where $u_T$ is the unitary component of the time-reversal operator ${\cal T}={\cal K}u_T^\dagger$ with complex conjugation ${\cal K}$. Fermi statistics further implies $\Delta(k)=-\Delta(-k)^T$. To determine the pairing state  we expand 
\begin{equation}
\Phi\left(k\right)=\sum_{J=1}^{n_{\gamma}^{2}}\chi_{J}\left(k\right)\Gamma_{J},   
\end{equation}
in a complete set of Hermitian $n_{\gamma}\times n_{\gamma}$ matrices $\Gamma_{J}$. Only solutions that are isotropic in momentum-frequency space and hence only a function of $|k|=\sqrt{\omega^2+\boldsymbol{k}^2}$ occur, where we set $v_{\rm F}=1$. The detailed $|k|$-dependence of $\chi_J(k)$ was analyzed in Ref.~\cite{Stangier2025}, but will not be important for our considerations,  except that  $\Phi\left(k\rightarrow 0\right)\neq 0$ without fine tuning. We focus on the structure of superconductivity in spinor space.  The superconducting
instability only occurs for sufficiently large dimensionless pairing strength
$\lambda_{p}$ such that the normal-state anomalous dimension exceeds a critical value,
i.e., $\eta_\psi \geq \eta_\psi^c$~\footnote{while $\eta_{\psi}^c\approx 0.14628$ within our theory~\cite{Stangier2025}, we regard the {\em existence\/} of such a minimum anomalous fermion dimension (rather than its precise value, which will depend on details beyond the scope of our analysis) as the key prediction of our theory.}.
Thus, within our theory
the {\em incoherent} nature of electrons at criticality is essential for, and drives, superconductivity.


We now turn to the question of the $\Gamma_J$ that contribute to $\Phi$, which 
establish the types of pairing order that can occur.  Indeed, for the dominant pairing instability of
Eq.~(\ref{Eq:linearizedEquation})
this is fully determined by rather easy-to-analyze algebraic conditions~\cite{Stangier2025}: It must hold that 
\begin{eqnarray}
  \left[\Gamma_{J},\alpha_{i}\right]&=&0 , \nonumber \\
  \Gamma_{J}\Upsilon_{a}&=&\tau_\phi \Upsilon_{a}\Gamma_{J},
  \label{eq:conditions}
\end{eqnarray}
for $i=1,2$ and  $a=1\cdots m_{\phi}$. 
In addition, the pairing state
must  obey Fermi statistics, i.e. $(\Gamma_{J}u_{T})^T=-\Gamma_{J}u_{T}$. If these conditions are fulfilled, superconductivity at the GN critical point will be possible.
One way to interpret these conditions is that possible pairings only come from the first condition along with the Pauli condition, i.e.  the possible pairings are independent of the interactions. We then use the second condition to see what couplings  can create them.  Assuming, for the moment, constant-in-$k$ mean-field order parameters for superconductivity and the critical boson, the conditions of Eq.~\eqref{eq:conditions} imply that isotropic gaps form in the spectrum of the corresponding Bogoliubov-de Gennes Hamiltonian, where all  gaps add up in squares; 
see appendix~\ref{appendixA}.
We will next scrutinize the possible pairing states and couplings compatible with Eq.~\eqref{eq:conditions} for twisted double-bilayer WSe$_{2}$ and TBG  within their respective Dirac theories.

\emph{Twisted double-bilayer WSe$_{2}$: }Recent experiments~\citep{Ma2024} identified
a semi-metal insulator transition as a function of the twist angle in AB-BA stacked twisted double-bilayer
WSe$_{2}$ which is considered a strongly-correlated version of single-layer
graphene~\citep{Angeli2021,Pan2023,Foutty2023,Ma2024,Tolosa2025} (and therefore a likely  candidate for
the present scenario).  To apply Eq.~(\ref{eq:conditions}) we need to
identify the Dirac matrices $\alpha_i$ and the coupling matrices $\Upsilon_a$.
The relevant quantum numbers are
spin, valley, and sub-lattice which we denote by the Pauli matrices $\sigma_{j_{1}}$, $\tau_{j_{2}}$, and $\rho_{j_{3}}$, respectively.
The $8\times 8$ matrices $\alpha_i$ are then given by:
\begin{equation}
    \left(\alpha_{1},\alpha_{2}\right)=\sigma_{0}\left(\tau_{3}\rho_{1},\tau_{0}\rho_{2}\right).
\end{equation}
 Time-reversal is determined by $u_{T}=i\sigma_{2}\tau_{1}\rho_{0}$. The complete set of matrices needed for the determination of the pairing state is most conveniently written as $\Gamma_{J}=\sigma_{j_{1}}\tau_{j_{2}}\rho_{j_{3}}$,
labeled by $J=\left(j_{1},j_{2},j_{3}\right)=1\cdots 64$.

Let us first discuss the possible collective bosons that induce upon condensation  a gap in the Dirac spectrum and whose fluctuations may serve as pairing glue. The corresponding  ordered states have all been discussed in great detail, see e.g. Refs.~\citep{Herbut2006,Herbut2009, Herbut2009b, Jurivcic2009, Weeks2010,Semenoff2012,Assaad2013}. 
Spin-orbit interaction in this system is believed to be very small~\citep{Angeli2021,Pan2023},
so it is natural to consider Heisenberg couplings of the type 
\begin{equation}
\vec{\Upsilon}_{{j_{2}},{j_{3}}}=(\Upsilon_{1,{j_{2}},{j_{3}}},\Upsilon_{2,{j_{2}},{j_{3}}},\Upsilon_{3,{j_{2}},{j_{3}}})=\vec{\sigma}\tau_{j_{2}}\rho_{j_{3}},
\end{equation}
or coupling in the charge sector
\begin{equation}
\Upsilon_{0,{j_{2}},{j_{3}}}=\sigma_{0}\tau_{j_{2}}\rho_{j_{3}}.
\end{equation}
For the combined valley and sub-lattice index $({j_{2}},{j_{3}})$ there are then
in total four options that open an isotropic gap, i.e. where the coupling
matrices anti-commute with $\alpha_{1}$ and $\alpha_{2}$ simultaneously.
This is the case for $\left({j_{2}},{j_{3}}\right)=\left(0,3\right)$, $\left(1,1\right)$,
$\left(2,1\right)$, and $(3,3)$. For the point group $D_{6}$ the
spatial parts of these interactions transform as ${\rm B}_{2}$, ${\rm A}_{1}$,
${\rm B}_{1}$, and ${\rm A}_{2}$, respectively. 
The first option
$\Upsilon_{\mu,03}=\sigma_{\mu}\tau_{0}\rho_{3}$ breaks the sub-lattice
symmetry (either in the spin or in the charge sector). $\left(\Upsilon_{\mu,11,}\Upsilon_{\mu,21}\right)=\sigma_{\mu}\left(\tau_{1},\tau_{2}\right)\rho_{1}$
are valley-mixing bond-order states akin to Kekul{\'e} order. They are closely related to
the inter-valley coherent (IVC) insulators that will be discussed in the
context of TBG below. Finally, $\Upsilon_{\mu,33}=\sigma_{\mu}\tau_{3}\rho_{3}$
corresponds for $\mu=0$ to the Haldane state~\cite{Haldane1988}, a loop-current ordered
state that breaks time-reversal symmetry, while $\vec{\Upsilon}_{33}=\vec{\sigma}\tau_{3}\rho_{3}$
is a time-reversal even ordered state  with opposite currents for the two spin flavors (parallel and anti-parallel to $\vec{\Upsilon}_{33}$), leading to an anomalous spin-Hall effect. Notice $\sigma_{3}\tau_{3}\rho_{3}$
is the celebrated Kane-Mele spin-orbit term~\cite{Kane2005} that would be spontaneously
generated in the gapped phase.

\begin{figure}
    \centering
    \includegraphics[width=0.95\linewidth]{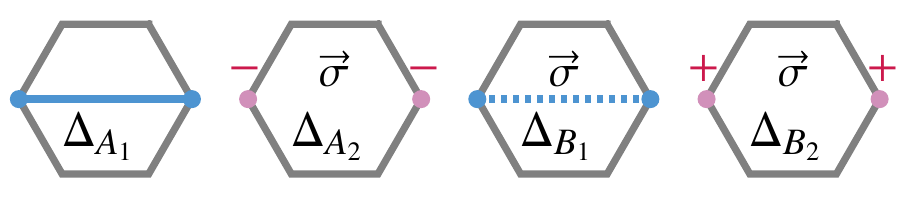}
    \\
    \includegraphics[width=0.95\linewidth]{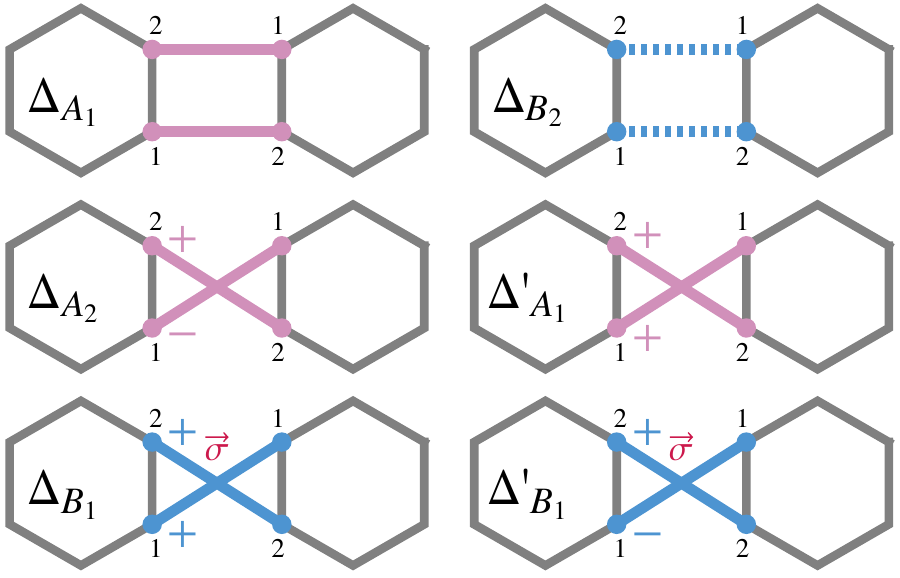}
    \caption{Top Row: The four possible SC instabilities in double-bilayer WSe$_2$ 
    given in Eqs.~(\ref{eq:pairing WSe2}), (\ref{eq:pairing WSe2-1-1}) and (\ref{eq:pairing WSe2-1}) labeled
    by the D$_6$ irrep. Pairing correlations that are intra-sublattice ($\propto \rho_0$)
    or inter-sublattice ($\propto \rho_2$) are labeled blue and pink, respectively, with
    solid (dashed) lines connecting inter-valley pairing of type $\tau_1$ ($\tau_2$).  
    Panels with no lines indicate intravalley pairing that is either in phase
    ($\propto \tau_0$, indicated with $+$) or out of phase  
    ($\propto \tau_3$, indicated with $-$). Bottom Rows:
    The six possible SC instabilities in TBG depending on the nature of the insulating state (KIVC or TIVC), as listed in Eqs.~(\ref{eq:pairing_KIVC}) and (\ref{eq:pairing_TIVC}) and labeled by the D$_6$ irrep. In each panel the left and right hexagons are the mini-BZ's associated with opposite valleys, with
    numbers $1,2$ labeling mini-valleys.  The valley pairing is  labeled by color (pink for $\tau_1$ and blue for $\tau_2$).  The pairing can also
    mix mini-valleys, with $\mu_1$ (or $\mu_2$) pairing given by solid (or dashed) lines, or be in the same 
    mini-valley, with $+-$ (or $++$) labeling $\mu_z$ (or $\mu_0$) giving the phase.
    In each case, $\vec{\sigma}$ 
    indicates triplet pairing, with the other cases singlet.
   }
    \label{fig:TWO}
\end{figure}


We have analyzed all possible pairing states due to fluctuations of
these collective bosons by analyzing the algebraic rules of Eq.~\eqref{eq:conditions} and  find  no stable superconducting
solution due to time-reversal odd  fluctuations $\vec{\Upsilon}_{03}$,
$\vec{\Upsilon}_{11}$, and $\vec{\Upsilon}_{21}$.
Equally, no superconductivity emerges due to fluctuations of the Haldane
state $\Upsilon_{0,33}$, also time-reversal odd. All other, time-reversal even states do, however, induce pairing: Charge-sub-lattice fluctuations with $\Upsilon_{0,03}$ induce $\vec{\Phi}=\vec{\sigma}\tau_{3}\rho_{0}$
which yields for the superconducting order parameter
\begin{equation}
\vec{\Delta}_{B_{1}}=i\sigma_{2}\vec{\sigma}\tau_{2}\rho_{0},\label{eq:pairing WSe2}
\end{equation}
i.e. a spin triplet, valley singlet. ${\rm B}_{1}$  gives the transformation  of the spatial part w.r.t.~$D_6$. The same coupling also induces a  singlet s-wave $\Phi=\sigma_{0}\tau_{0}\rho_{0}$, which becomes 
\begin{equation}
\Delta_{A_{1}}=i\sigma_{2}\tau_{1}\rho_{0}.\label{eq:pairing WSe2-1-1}
\end{equation}
$\vec{\Delta}_{B_{1}}$ and $\Delta_{A_{1}}$ are, at the
level of the continuum theory, degenerate; a degeneracy
that will  be lifted by lattice corrections to the Dirac theory. 

Next, we
consider valley-mixing Kekul{\'e}-type bond order fluctuation $\Upsilon_{0,11}$
and $\Upsilon_{0,21}$. If we consider them individually, as single-component
order parameters, they induce either  $\vec{\Phi}=\vec{\sigma}\tau_{2}\rho_{2}$ or
$\vec{\Phi}=\vec{\sigma}\tau_{1}\rho_{2}$, which correspond to 
\begin{equation}
\vec{\Delta}_{{\rm A}_{2}}=i\sigma_{2}\vec{\sigma}\tau_{3}\rho_{2}\,\,{\rm and}\,\,\vec{\Delta}_{{\rm B}_{2}}=i\sigma_{2}\vec{\sigma}\tau_{0}\rho_{2}.\label{eq:pairing WSe2-1}
\end{equation}
These are spin triplets and sub-lattice singlets with out-of-phase
or in-phase order parameter in the valleys. They  are again
degenerate with the $s$-wave of Eq.~\eqref{eq:pairing WSe2-1-1}.
Alternatively, we can consider $\left(\Upsilon_{0,11},\Upsilon_{0,21}\right)$
as a degenerate two-component order parameter, a degeneracy caused by the $U_{\rm valley}(1)$ symmetry that, within the Dirac theory, conserves  charge in each valley separately. Then both spin triplets are
frustrated by the other boson component, that acts pair-breaking. As a consequence, only  $\Delta_{A_{1}}$ survives.
Finally, the anomalous spin-Hall state $\vec{\Upsilon}_{33}=\vec{\sigma}\tau_{3}\rho_{3}$
 gives  rise to the s-wave state $\Delta_{A_{1}}$ of Eq.~\eqref{eq:pairing WSe2-1-1}.
The four possible pairing states of WSe$_2$ are
illustrated in Fig.~\ref{fig:TWO} (top row).
 
To summarize,  superconductivity at the critical twist angle for twisted double-bilayer WSe$_2$  occurs for bosons that do not break time-reversal symmetry. The most frequent pairing state is the $s$-wave of Eq.~\eqref{eq:pairing WSe2-1-1}.

\emph{Twisted bilayer graphene: } In this case we need to include, in addition to spin, valley and sub-lattice, the mini-valley quantum number  (the two points $\boldsymbol{K}$
and $\boldsymbol{K}'$ of the mini Brillouin zone)~\citep{Po2018,Kang2019,Bultinck2020,Christos2020,Parthenios2023}, which we denote by Pauli matrices $\mu_{j_{4}}$
Hence, we  now have $n_{\gamma}=16$-component
Dirac spinors and will expand our pairing states in the $256$ matrices $\Gamma_{J}=\sigma_{j_{1}}\tau_{j_{2}}\rho_{j_{3}}\mu_{j_{4}}$ and analyze the conditions in Eq.~\eqref{eq:conditions}  to determine which of them is an allowed 
superconducting
state.  The Dirac dispersion is described by the two Dirac matrices 
\begin{equation}
    \left(\alpha_{1},\alpha_{2}\right)=\sigma_{0}\left(\tau_{3}\rho_{1}, \tau_{0}\rho_{2}\right)\mu_{0}.
\end{equation}
Time reversal is determined by $u_{T}=i\sigma_{2}\tau_{1}\rho_{0}\mu_{0}$. For 
simplicity
we  focus on bosons that yield 
IVC
insulators, i.e.~that break the translation invariance of the original graphene lattice by forming bond- or loop-current order~\cite{Kang2019,Bultinck2020} and that have been observed, albeit at different fillings~\cite{Nuckolls2023}. The effects of all other gap-inducing bosons for TBG are listed in appendix~\ref{appendixA}. 
The coupling of
 Dirac fermions to an IVC state depends on whether it does or does not break time-reversal symmetry~\citep{Ingham2023,Huang2025}. In the former case we have the Kramers-IVC (KIVC) with
coupling matrices
\begin{eqnarray}
\left(\Upsilon_{1},\Upsilon_{2}\right)_{{\rm KIVC}} & = & \sigma_{0}\left(\tau_{1},\tau_{2}\right)\rho_{1}\mu_{2},
\end{eqnarray}
or, alternatively, the time-reversal symmetric
IVC (TIVC) with
\begin{eqnarray}
\left(\Upsilon_{1},\Upsilon_{2}\right)_{{\rm TIVC}} & = & \sigma_{0}\left(\tau_{1},\tau_{2}\right)\rho_{1}\mu_{1}.
\end{eqnarray}
In appendix~\ref{appendixA} we discuss similar results for other TIVC states.
In both cases we consider a two-component or $XY$-order parameter. 

For both interactions we find from the algebraic conditions, Eq.~\eqref{eq:conditions} four  degenerate pairing solutions. In the case of KIVC they are
\begin{eqnarray}
\Delta_{{\rm A}_{1}} & \propto & i\sigma_{2}\tau_{1}\rho_{0}\mu_{1},\nonumber \\
\Delta_{{\rm A}_{2}} & \propto & i\sigma_{2}\tau_{1}\rho_{0}\mu_{3},\nonumber \\
\vec{\Delta}_{{\rm B}_{1}} & \propto & i\sigma_{2}\vec{\sigma}\tau_{2}\rho_{0}\mu_{0},\nonumber \\
\Delta_{{\rm B}_{2}} & \propto & i\sigma_{2}\tau_{2}\rho_{0}\mu_{2}.\label{eq:pairing_KIVC}
\end{eqnarray}
Except for the spin triplet B$_1$ state, all of these superconductors are spin singlets. While the A$_1$ order parameter transforms trivially under all point group symmetries, it corresponds to a finite-momentum superconductor with non-trivial phase shift under moir\'e translation; this is also true for B$_2$ in Eq.~(\ref{eq:pairing_KIVC}) 
while
the remaining two states have Cooper pairs with zero center-of-mass momentum.


Considering TIVC fluctuations,
$\Delta{}_{{\rm A}_{1}}$ and $\Delta{}_{{\rm B}_{2}}$
are unchanged while instead of the other two, we now find
\begin{eqnarray}
\Delta'_{{\rm A}_{1}} & \propto & i\sigma_{2}\tau_{1}\rho_{0}\mu_{0},\nonumber \\
\vec{\Delta}'_{{\rm B}_{1}} & \propto & i\sigma_{2}\vec{\sigma}\tau_{2}\rho_{0}\mu_{3}.
\label{eq:pairing_TIVC}
\end{eqnarray}
These four states are again degenerate at the level of the Dirac theory.  
The six distinct  pairing states of TBG, discussed above, are
illustrated in Fig.~\ref{fig:TWO} (bottom three rows).
Most importantly, in TBG we find superconductivity emerging at the onset of inter-valley coherent insulators, irrespective of whether time-reversal symmetry is broken or preserved. The additional mini-valley flavor structure substantially enhances the range of possible pairing states, in comparison to WSe$_2$.

To estimate the 
superconducting transition temperature
$T_{{\rm c}}$ in these twisted Dirac 
materials, we note that the maximal 
$T_{{\rm c}}$ 
obtained in Ref.~\cite{Stangier2025} is approximately $2\times10^{-3}\Lambda$, where $\Lambda$ denotes
the cutoff energy of the Dirac model. For both WSe$_{2}$ at a twist
angle of $\theta\approx2.7^\circ$ and for TBG at
$\theta\approx1.2^\circ$ one finds $\Lambda\sim20\,{\rm meV}$~\cite{Ma2024,Xiao2025},
which yields an optimal transition temperature of $T_{{\rm c}}\approx0.4\,{\rm K}$.

For both systems, our results for pairing at the semi-metal to insulator transition were obtained in the  limit where we only consider the Dirac point and ignore the behavior in the rest of the mini-Brillouin zone or due to higher bands. Hence, we should ask:
Are there implications beyond 
the Dirac physics near the Gross-Neveu critical point?  In this context, it is interesting that the second condition in Eq.~\eqref{eq:conditions} was recently obtained
in the study of pairing instabilities of TBG in the extreme flat-band limit~\cite{Christos2023nodal}, i.e. not in the Dirac regime. 
This strongly suggests that it may in fact be of more general relevance than either limit. Of course, the first condition in Eq.~\eqref{eq:conditions} is specific to
the Dirac theory.

Of the many approaches to link superconductivity to nearby ordered states, a particularly beautiful and powerful one is the observation that the seeds of pairing are planted in the topological excitations of the insulator and vice versa~\cite{Wiegmann1999,Abanov2000theta,Grover2008,Christos2020,Khalaf2021charged,Ledwith2021strong}. This is the result of a Wess-Zumino-Witten term 
\begin{equation}
    S_{{\rm WZW}}=i\frac{3{\cal N}}{4\pi}\int \epsilon_{\alpha\beta\gamma\delta\nu}\, n_{\alpha}\partial_{u}n_{\beta}\partial_{\tau}n_{\gamma}\partial_{x}n_{\delta}\partial_{y}n_{\nu},
    \label{eq:SWZW}
\end{equation}
of the coupled superconducting and mass-generating boson problem. The five-component order parameter $n_\alpha$ comprises the two components of the complex superconducting order parameter together with three suitably chosen collective bosonic modes, denoted by $\vec{\Upsilon}$. The integration
$\int \cdots = \int d^{3}x \int_{0}^{1} du$
extends over space-time and the auxiliary variable $u$~\cite{Wiegmann1999,Abanov2000theta,Grover2008,Christos2020,Khalaf2021charged,Ledwith2021strong}. Ref.~\cite{Christos2020} formulated three conditions for ${\cal N}\neq 0$ in a particularly convenient way for our purposes. Remarkably, the pairing conditions of Eq.~\eqref{eq:conditions}, derived from the analysis of the critical pairing problem, coincide with two of these three criteria. The remaining condition imposes an additional constraint on the three components of $\vec{\Upsilon}$, namely
\begin{equation}
{\rm tr\left(\alpha_{i_{1}}\alpha_{i_{2}}\Upsilon_{a_{1}}\Upsilon_{a_{2}}\Upsilon_{a_{3}}\right)} =-8{\cal N} \epsilon_{i_{1}i_{2}a_{1}a_{2}a_{3}}\neq 0,
\label{eq:conditions_WZW_2}
\end{equation}
which also determines the coefficient $\cal N$ in Eq.~\eqref{eq:SWZW}.

Analyzing options for such topologically enabled pairing for WSe$_2$,  we find the partner states $\vec{\Upsilon}=\left(\Upsilon_{011,}\Upsilon_{021},\Upsilon_{003}\right)$  and  $\vec{\Upsilon}_{33}$, both with $\Delta_{A_{1}}$ of Eq.~\eqref{eq:pairing WSe2-1-1}. 
Hence, either a combination of bond with charge order or the anomalous spin-Hall state, already discussed in Ref.~\cite{Grover2008}, form natural partner orders with $s$-wave superconductivity to yield a topological origin of pairing. Here ${\cal N}=1$ and skyrmion defects in the insulator carry charge $2e$.
For the TBG problem, the behavior is again much richer and we find in total $56$ such partner states, all combined with a wide array
of superconducting states. They are listed in appendix~\ref{appendixB}, see also Ref.~\cite{Christos2020}.
Examples are the two states for WSe$_2$ multiplied by the unit matrix in mini-valley space: $\vec{\Upsilon}\mu_0$ and $\vec{\Upsilon}_{33}\mu_0$. As a result of the additional mini-valley quantum number, we have ${\cal N}=2$ so that skyrmions carry charge $4e$.
For both systems, we find the common theme that only bosons that are even under time-reversal yield non-zero $S_{{\rm WZW}}$.

These results are rather unexpected.  {\em A priori}, there is no reason to expect a connection between the conditions arising from a pairing instability due to critical boson exchange and those required for a topological term in the action. Yet our analysis implies that all insulators with charged skyrmions, together with their superconducting partners, form a subset of the states we obtain from Eq.~\eqref{eq:conditions}. Moreover, our theory shows not only that these particle-hole / particle-particle partners are natural candidates for topological pairing, but also that the superconducting partners are in fact induced by fluctuations of  particle-hole modes. Consequently, the pairing states obtained here may be of significance even beyond neutrality or the specific critical point analyzed.

In summary, we investigate superconductivity emerging near relativistic
Mott transitions in twisted two-dimensional Dirac materials. Using
a controlled strong-coupling framework, based on a generalized Sachdev-Ye-Kitaev
approach,  it was shown in Ref.~\cite{Stangier2025} that superconductivity may appear once the fermionic anomalous dimension induced by Gross\textendash Neveu criticality
exceeds a universal threshold, enabling attractive pairing to overcome
the bare Dirac semi-metal behavior. Applying this theory to twisted double-bilayer WSe$_{2}$, we find that all time-reversal even bosons cause superconductivity.
In contrast, for TBG we find a manifold of degenerate and nearly degenerate
superconducting states that can be either time-reversal even or odd. Broadly speaking, the more intricate the Dirac structure (i.e., the larger the Dirac representation of the twisted material), the more readily pairs can form. 
Finally, we established a direct link between pairing due to critical bosons and the formation of charged topological solitons, i.e. skyrmions, in the associated insulator. The latter are a subset of the former. Twisted Dirac materials provide a controlled and tunable setting to probe the interplay between relativistic Mott criticality, the associated insulating states, and unconventional superconductivity.

We are grateful to  Andrey V. Chubukov,  Laura Classen,  Elio K{\"o}nig,  and Nikolaos Parthenios for helpful discussions. This work was supported
by the German Research Foundation TRR 288-422213477 ELASTO-Q-MAT,
 B01 (V.C.S. and J.S.) and grant  SFI-MPS-
NFS-00006741-05 from the Simons Foundation (J.S.).  
D.E.S. acknowledges support from the National Science Foundation under Grant PHY-2208036. 
M.S.S.~acknowledges funding by the European Union (ERC-2021-STG, Project 101040651---SuperCorr). Views and opinions expressed are however those of the authors only and do not necessarily reflect those of the European Union or the European Research Council Executive Agency. Neither the European Union nor the granting authority can be held responsible for them.

\bibliography{refs}

\clearpage
\appendix

\title{Supplementary Material to: "Superconductivity of Incoherent Electrons  near the Relativistic Mott Transition in Twisted Dirac Materials" }
\author{Veronika C. Stangier }
\affiliation{Institute for Theory of Condensed Matter, Karlsruhe Institute of Technology,
Karlsruhe 76131, Germany}
\author{Mathias S. Scheurer}
\affiliation{Institute for Theoretical Physics III, University of Stuttgart,  Stuttgart 70550, Germany}
\author{Daniel E. Sheehy }
\affiliation{Department of Physics and Astronomy, Louisiana State University, the
Baton Rouge, LA 70803 USA}
\author{J{\"o}rg Schmalian }
\affiliation{Institute for Theory of Condensed Matter, Karlsruhe Institute of Technology,
Karlsruhe 76131, Germany}
\affiliation{Institute for Quantum Materials and Technologies, Karlsruhe Institute
of Technology, Karlsruhe 76131, Germany}
\date{\today }
\maketitle
\section{Interpretation of the pairing conditions and lists of pairing states}
\label{appendixA}
To interpret the conditions in Eq.~\eqref{eq:conditions} of the main paper physically, we assume for simplicity
$\Phi\left(k\right)=\chi\Gamma_{J}$ with $\chi$ a constant function
of $k$ and consider  finite expectation values $\mu_{a}\sim\left\langle \phi_{a}\right\rangle $. The normal state Hamiltonian becomes $h_{\boldsymbol{k}}=v_{{\rm F}}\boldsymbol{k}\cdot\boldsymbol{\alpha}+\mu_{a}\Upsilon_{a}$ which yields  the  Bogoliubov-de Gennes Hamiltonian
\begin{equation}
{\cal H}_{\boldsymbol{k}}^{{\rm BdG}}=\left(\begin{array}{cc}
h_{\boldsymbol{k}} & \Phi_{\boldsymbol{k}}\\
\Phi_{\boldsymbol{k}}^{\dagger} & -u_{T}^{\dagger}h_{\boldsymbol{-k}}^{*}u_{T}
\end{array}\right)=v_{{\rm F}}\boldsymbol{k}\cdot\tilde{\boldsymbol{\alpha}}+\sum_{r=1}^{m_{\phi}+2}\mu_{r}\tilde{\beta}_{r},
\end{equation}
with $\tilde{\alpha}_{i}=\alpha_{i}\kappa_{z}$ and additional Pauli
matrices $\kappa_{a}$ in Nambu space. The masses $\mu_{r}=\left(\mu_{1},\cdots,\mu_{m_{\phi}},{\rm Re}\chi,-{\rm Im\chi}\right)$
are either due to the condensation of the bosonic mode or the superconducting
order parameter. For $1\leq r\leq m_{m_{\phi}}$ holds $\tilde{\beta}_{r}=\Upsilon_{r}\kappa_{z}$
if $\tau_{\phi}=1$ or $\tilde{\beta}_{r}=\Upsilon_{r}\kappa_{0}$
if $\tau_{\phi}=-1$. In addition $\tilde{\beta}_{r=m_{\phi}+1}=\Gamma_{J}\kappa_{x}$
and $\tilde{\beta}_{r=m_{\phi}+2}=\Gamma_{J}\kappa_{y}$ describe
the superconducting gap. The conditions, Eq.~(5) then become $\left\{ \tilde{\beta}_{r},\tilde{\beta}_{r'}\right\} =\delta_{r,r'}$
and $\left\{ \tilde{\alpha}_{i},\tilde{\beta}_{r'}\right\} =0$. Hence
all mass terms add in squares with Bogoliubov energies 
$E_{\boldsymbol{k}}=\pm\sqrt{(v_{{\rm F}}\boldsymbol{k})^{2}+\sum_{r}\mu_{r}^{2}}$.

Finally, we list all pairing states for the
Dirac theories of twisted double bilayer WSe$_{2}$ and twisted bilayer
graphene. The algebraic conditions for pairing are given in Eq.~\eqref{eq:conditions}.
The pairing states for AB-BA stacked twisted double-bilayer WSe$_{2}$ at filling $\nu=2$ are listed in Table \ref{tab=00007BTableWSe2}, while the states for twisted bilayer graphene  are listed in Table \ref{tab:TableTBG}.
\begin{table*}
\begin{tabular}{|c|c|c|c|c|}
\hline 
$\Upsilon_{i}$ & ${\cal T}(\sigma_{0})$ & ${\rm D}_{6}$ & $\Phi$ for $\sigma_{0}$ in $\Upsilon_{i}$ & $\Phi$ for $\boldsymbol{\sigma}$ in $\Upsilon_{i}$\tabularnewline
\hline 
\hline 
$\sigma_{\mu}\left(\tau_{1},\tau_{2}\right)\rho_{1}$ & $+$ & $\left({\rm A}_{1},{\rm B}_{1}\right)$ & $\sigma_{0}\tau_{0}\rho_{0}$ & -\tabularnewline
\hline 
$\sigma_{\mu}\tau_{0}\rho_{3}$ & $+$ & ${\rm B}_{2}$ & $\sigma_{0}\tau_{0}\rho_{0}$, $\boldsymbol{\sigma}\tau_{3}\rho_{0}$ & -\tabularnewline
\hline 
$\sigma_{\mu}\tau_{3}\rho_{3}$ & $-$ & ${\rm A}_{2}$ & - & $\sigma_{0}\tau_{0}\rho_{0}$\tabularnewline
\hline 
\end{tabular}

\caption{Coupling matrices for WSe$_{2}$ that anti-commute with the $\alpha_{1,2}$,
along with their transformation under time-reversal (we list the charge
sector ($\sigma_0$), the spin sector ($\boldsymbol{\sigma}$) transforms oppositely) and the transformation
under ${\rm D}_{6}$. The pairing states for the charge and spin sector
bosons are listed in the last two columns. Only time-reversal even
interactions cause superconductivity. Two TIVC states in the first
row are grouped into a pair that transforms into each other under
$U_{{\rm v}}\left(1\right)=\exp\left(i\zeta\tau_{3}\right)$, i.e.
they approximately behave like $XY$ degrees of freedom. Individually,
each would also induce another triplet state (not listed), for which
the other $XY$ component is however repulsive.}
\label{tab=00007BTableWSe2}
\end{table*}

\begin{table*}
\begin{tabular}{|c|c|c|c|c|}
\hline 
$\Upsilon_{i}$ & ${\cal T}(\sigma_{0})$ & ${\rm D}_{6}$ & $\Phi$ for $\sigma_{0}$ in $\Upsilon_{i}$ & $\Phi$ for $\boldsymbol{\sigma}$ in $\Upsilon_{i}$\tabularnewline
\hline 
\hline 
$\sigma_{\mu}\left(\tau_{1},\tau_{2}\right)\rho_{1}\mu_{1}$ & $+$ & $\left({\rm A}_{1},{\rm B}_{1}\right)$ & %
\begin{tabular}{cc}
$\sigma_{0}\tau_{0}\rho_{0}\mu_{0,1}$, & $\sigma_{0}\tau_{3}\rho_{0}\mu_{2}$,\tabularnewline
$\boldsymbol{\sigma}\tau_{3}\rho_{0}\mu_{3}$ & \tabularnewline
\end{tabular} & $\sigma_{0}\tau_{0}\rho_{0}\mu_{3}$\tabularnewline
\hline 
$\sigma_{\mu}\left(\tau_{1},\tau_{2}\right)\rho_{1}\mu_{2}$ & $-$ & $\left({\rm A}_{2},{\rm B}_{2}\right)$ & %
\begin{tabular}{cc}
$\sigma_{0}\tau_{0}\rho_{0}\mu_{1,3}$, & $\sigma_{0}\tau_{3}\rho_{0}\mu_{2}$,\tabularnewline
$\boldsymbol{\sigma}\tau_{3}\rho_{0}\mu_{0}$ & \tabularnewline
\end{tabular} & $\sigma_{0}\tau_{0}\rho_{0}\mu_{0}$\tabularnewline
\hline
$\sigma_{\mu}\left(\tau_{1},\tau_{2}\right)\rho_{1}\mu_{3}$ & $+$ & $\left({\rm A}_{2},{\rm B}_{2}\right)$ & %
\begin{tabular}{cc}
$\sigma_{0}\tau_{0}\rho_{0}\mu_{0,3}$, & $\sigma_{0}\tau_{3}\rho_{0}\mu_{2}$,\tabularnewline
$\boldsymbol{\sigma}\tau_{3}\rho_{0}\mu_{1}$ & \tabularnewline
\end{tabular} & $\sigma_{0}\tau_{0}\rho_{0}\mu_{1}$ \\[-1.1ex]  
\multicolumn{5}{c}{\hspace{-0.4em}\hdashrule[0.5ex]{\dimexpr0.7\linewidth-2\tabcolsep\relax}{0.4pt}{1mm 1mm}} \hspace{-1.2em} \\[-1.2ex] 
$\sigma_{\mu}\left(\tau_{1},\tau_{2}\right)\rho_{1}\mu_{0}$ & $+$ & $\left({\rm A}_{1},{\rm B}_{1}\right)$ & $\sigma_{0}\tau_{0}\rho_{0}\mu_{0,1,3}$, \textbf{$\boldsymbol{\sigma}\tau_{0}\rho_{0}\mu_{2}$} & $\sigma_{0}\tau_{3}\rho_{0}\mu_{2}$\tabularnewline
\hline  
$\sigma_{\mu}\tau_{0}\rho_{3}\mu_{0}$ & $+$ & ${\rm B}_{2}$ & %
\begin{tabular}{cc}
$\sigma_{0}\tau_{0}\rho_{0}\mu_{0,1,3}$, & $\sigma_{0}\tau_{3}\rho_{0}\mu_{2}$,\tabularnewline
$\boldsymbol{\sigma}\tau_{3}\rho_{0}\mu_{0,1,3}$, & $\boldsymbol{\sigma}\tau_{0}\rho_{0}\mu_{2}$\tabularnewline
\end{tabular} & $\sigma_{0}\tau_{1,2}\rho_{2}\mu_{2}$\tabularnewline
\hline 
$\sigma_{\mu}\tau_{3}\rho_{3}\mu_{0}$ & $-$ & ${\rm A}_{2}$ & $-$ & %
\begin{tabular}{cc}
$\sigma_{0}\tau_{0}\rho_{0}\mu_{0,1,3}$, & $\sigma_{0}\tau_{3}\rho_{0}\mu_{2}$,\tabularnewline
$\sigma_{0}\tau_{1,2}\rho_{2}\mu_{2}$ & \tabularnewline
\end{tabular}\tabularnewline
\hline 
$\sigma_{\mu}\tau_{0}\rho_{3}\mu_{1}$ & $+$ & ${\rm B}_{2}$ & %
\begin{tabular}{cc}
$\sigma_{0}\tau_{0}\rho_{0}\mu_{0,1}$, & $\sigma_{0}\tau_{1,2}\rho_{2}\mu_{2}$,\tabularnewline
$\boldsymbol{\sigma}\tau_{3}\rho_{0}\mu_{0,1}$, & $\boldsymbol{\sigma}\tau_{1,2}\rho_{2}\mu_{3}$\tabularnewline
\end{tabular} & $\sigma_{0}\tau_{3}\rho_{0}\mu_{2}$, $\sigma_{0}\tau_{0}\rho_{0}\mu_{3}$\\[-1.1ex]  
\multicolumn{5}{c}{\hspace{-0.4em}\hdashrule[0.5ex]{\dimexpr0.7\linewidth-2\tabcolsep\relax}{0.4pt}{1mm 1mm}} \hspace{-1.2em} \\[-1.2ex]
$\sigma_{\mu}\tau_{3}\rho_{3}\mu_{2}$ & $+$ & ${\rm A}_{1}$ & \multicolumn{1}{c|}{%
\begin{tabular}{cc}
$\sigma_{0}\tau_{0}\rho_{0}\mu_{0}$, & $\text{\ensuremath{\sigma_{0}\tau_{3}\rho_{0}\mu_{2}}}$,\tabularnewline
$\sigma_{0}\tau_{1,2}\rho_{2}\mu_{2}$, & $\boldsymbol{\sigma}\tau_{0}\rho_{0}\mu_{2}$,\tabularnewline
$\boldsymbol{\sigma}\tau_{3}\rho_{0}\mu_{0}$, & $\boldsymbol{\sigma}\tau_{1,2}\rho_{2}\mu_{0}$\tabularnewline
\end{tabular}} & $\sigma_{0}\tau_{0}\rho_{0}\mu_{1,3}$\tabularnewline
\hline 
$\sigma_{\mu}\tau_{3}\rho_{3}\mu_{1}$ & $-$ & ${\rm A}_{2}$ & %
\begin{tabular}{cc}
$\sigma_{0}\tau_{3}\rho_{0}\mu_{2}$, & $\text{\ensuremath{\sigma_{0}\tau_{0}\rho_{0}\mu_{3}}}$,\tabularnewline
$\sigma_{0}\tau_{1,2}\rho_{2}\mu_{2}$, & $\boldsymbol{\sigma}\tau_{0}\rho_{0}\mu_{2},$\tabularnewline
$\boldsymbol{\sigma}\tau_{3}\rho_{0}\mu_{3}$, & $\boldsymbol{\sigma}\tau_{1,2}\rho_{2}\mu_{3}$\tabularnewline
\end{tabular} & $\sigma_{0}\tau_{0}\rho_{0}\mu_{0,1}$ \\[-1.1ex]  
\multicolumn{5}{c}{\hspace{-0.4em}\hdashrule[0.5ex]{\dimexpr0.7\linewidth-2\tabcolsep\relax}{0.4pt}{1mm 1mm}} \hspace{-1.2em} \\[-1.2ex] 
$\sigma_{\mu}\tau_{0}\rho_{3}\mu_{2}$ & $-$ & ${\rm B}_{1}$ & %
\begin{tabular}{cc}
$\sigma_{0}\tau_{0}\rho_{0}\mu_{1,3}$, & $\sigma_{0}\tau_{1,2}\rho_{2}\mu_{2}$,\tabularnewline
$\boldsymbol{\sigma}\tau_{1,2}\rho_{2}\mu_{0}$, & $\boldsymbol{\sigma}\tau_{3}\rho_{0}\mu_{1,3}$\tabularnewline
\end{tabular} & $\sigma_{0}\tau_{0}\rho_{0}\mu_{0}$, $\sigma_{0}\tau_{3}\rho_{0}\mu_{2}$\tabularnewline
\hline 
$\sigma_{\mu}\tau_{0}\rho_{3}\mu_{3}$ & + & ${\rm B}_{1}$ & %
\begin{tabular}{cc}
$\sigma_{0}\tau_{0}\rho_{0}\mu_{0,3}$, & $\sigma_{0}\tau_{1,2}\rho_{2}\mu_{2}$,\tabularnewline
$\boldsymbol{\sigma}\tau_{1,2}\rho_{2}\mu_{1}$, & $\boldsymbol{\sigma}\tau_{3}\rho_{0}\mu_{0,3}$\tabularnewline
\end{tabular} & $\sigma_{0}\tau_{0}\rho_{0}\mu_{1}$, $\sigma_{0}\tau_{3}\rho_{0}\mu_{2}$\tabularnewline
\hline 
$\sigma_{\mu}\tau_{3}\rho_{3}\mu_{3}$ & $-$ & ${\rm A}_{1}$ & %
\begin{tabular}{cc}
$\sigma_{0}\tau_{0}\rho_{0}\mu_{1}$, & $\text{\ensuremath{\sigma_{0}\tau_{3}\rho_{0}\mu_{2}}}$,\tabularnewline
$\sigma_{0}\tau_{1,2}\rho_{2}\mu_{2}$, & $\boldsymbol{\sigma}\tau_{3}\rho_{0}\mu_{1}$,\tabularnewline
$\boldsymbol{\sigma}\tau_{1,2}\rho_{2}\mu_{1}$, & $\boldsymbol{\sigma}\tau_{0}\rho_{0}\mu_{2}$\tabularnewline
\end{tabular} & $\sigma_{0}\tau_{0}\rho_{0}\mu_{0,3}$\tabularnewline
\hline 
\end{tabular}

\caption{Same as Table~\ref{tab=00007BTableWSe2} but for TBG.
The first two lines correspond to TIVC and KIVC, respectively. These IVCs are listed as two-component order parameters as they transform into each other under the broken $U_{{\rm v}}\left(1\right)=\exp\left(i\zeta\tau_{3}\right)$, i.e. they approximately behave like $XY$ degrees of freedom. For mini-valleys, intravalley (intervalley) states with $\mu_1,$ and $\mu_2$ ($\mu_{0}$ and $\mu_{3}$) can also be grouped together as they transform into each other under translational symmetry on the moir\'e scale, in analogy to the $\tau_{1,2}$ states and graphene-scale translational symmetry. These pairs are indicated by the dashed lines between them. If we assume they appear together, this would reduce the number of pairing states: only states  occur in both rows remain.}

\label{tab:TableTBG}
\end{table*}

\section{List of partner states}
\label{appendixB}
In this section we list all bosonic parts of the partner states that
give rise to a Wess-Zumino-Witten term $S$$_{{\rm WZW}}$ in the
action. Conditions for the presence of $S$$_{{\rm WZW}}$ of the
coupled superconducting and mass-generating boson problem were listed
in Ref.~\cite{Christos2020} (see Eqs.(14a)-(14c) of
Ref.~\cite{Christos2020}), which in our notation corresponds to:
\begin{eqnarray}
\alpha_{i}\Gamma_{J}u_{T}^{\dagger} & = & -\Gamma_{J}u_{T}^{\dagger}\alpha_{i}^{T}\neq0\,\,\,{\rm for}\,\,i=1,2\nonumber \\
\Upsilon_{l}\Gamma_{J}u_{T}^{\dagger} & = & \Gamma_{J}u_{T}^{\dagger}\Upsilon_{l}^{T}\neq0\,\,\,{\rm for}\,\,l=1,2,3\nonumber \\
{\rm tr\left(\alpha_{i_{1}}\alpha_{i_{2}}\Upsilon_{l_{1}}\Upsilon_{l_{2}}\Upsilon_{l_{3}}\right)} & \propto & \epsilon_{i_{1}i_{2}l_{1}l_{2}l_{3}}\neq0.\label{eq:WZW_cond}
\end{eqnarray}
With $u_{T}^{\dagger}\alpha_{i}^{T}u_{T}=-\alpha_{i}$, we rewrite
the first equation as $\alpha_{i}\Gamma_{J}=\Gamma_{J}\alpha_{i}$,
which is the first condition Eq.~\eqref{eq:conditions} of the main text. With $u_{T}^{\dagger}\Upsilon_{l}^{T}u_{T}=\tau\Upsilon_{l}$,
the second line in Eq.~(\ref{eq:WZW_cond}) becomes $\Upsilon_{l}\Gamma_{J}=\tau\Gamma_{J}\Upsilon_{l}$, which
is precisely the second condition in Eq.~\eqref{eq:conditions} of the main text. Hence,
terms that give rise to a WZW term are a subset of those that make
up the pairing states of our Dirac theory. To identify this subset,
we need to analyze the last condition in Eq.~(\ref{eq:WZW_cond}).

Performing this analysis for the Dirac theory of WSe$_{2}$ we find
$20$ triples that obey Eq.~(\ref{eq:WZW_cond}). Each of them gives rise to a nontrivial pairing
state. If we consider only those where all three components transform
the same under time reversal, there are only two such options left
and both are even under time reversal. They are the anomalous spin-Hall
state 
\begin{equation}
\vec{\Upsilon}_{33}=\boldsymbol{\sigma}\tau_{3}\rho_{3},
\end{equation}
and the combination 
\begin{equation}
\vec{\Upsilon}=\sigma_{0}\left(\tau_{1}\rho_{1},\tau_{2}\rho_{1},\tau_{0}\rho_{3}\right)
\end{equation}
of the TIVC $XY$-state and charge sub-lattice order. Both were also
listed in the main text. The superconducting partner state is $\Phi\sim e^{i\theta}\sigma_{0}\tau_{0}\rho_{0}$ 

Looking for triples of order parameters that obey Eq.~(\ref{eq:WZW_cond})
for the Dirac model of twisted bilayer graphene, we find in total
$336$ distinct triples. If we confine ourselves again to the ones
where all three components of the order parameter transform the same
under time reversal we find $56$ triples. They are also all even
under time reversal and given by 
\begin{eqnarray}
\vec{\Upsilon}_{{\rm 1}} & = & \sigma_{0}\left(\tau_{1}\rho_{1},\tau_{2}\rho_{1},\tau_{0}\rho_{3}\right)\mu_{0}\nonumber \\
\vec{\Upsilon}_{{\rm 2}} & = & \sigma_{0}\left(\tau_{1}\rho_{1}\mu_{0},\tau_{2}\rho_{1}\mu_{1},\tau_{0}\rho_{3}\mu_{1}\right)\nonumber \\
\vec{\Upsilon}_{{\rm 3-5}} & = & \left(\sigma_{0}\tau_{1}\rho_{1}\mu_{0},\sigma_{i}\tau_{2}\rho_{1}\mu_{2},\sigma_{i}\tau_{0}\rho_{3}\mu_{2}\right)\nonumber \\
\vec{\Upsilon}_{{\rm 6}} & = & \sigma_{0}\left(\tau_{1}\rho_{1}\mu_{0},\tau_{2}\rho_{1}\mu_{3},\tau_{0}\rho_{3}\mu_{3}\right)\nonumber \\
\vec{\Upsilon}_{{\rm 7}} & = & \sigma_{0}\left(\tau_{2}\rho_{1}\mu_{0},\tau_{1}\rho_{1}\mu_{1},\tau_{0}\rho_{3}\mu_{1}\right)\nonumber \\
\vec{\Upsilon}_{{ 8-10}} & = & \left(\sigma_{0}\tau_{2}\rho_{1}\mu_{0},\sigma_{i}\tau_{1}\rho_{1}\mu_{2},\sigma_{i}\tau_{0}\rho_{3}\mu_{2}\right)\nonumber \\
\vec{\Upsilon}_{{ 11}} & = & \sigma_{0}\left(\tau_{2}\rho_{1}\mu_{0},\tau_{1}\rho_{1}\mu_{3},\tau_{0}\rho_{3}\mu_{3}\right)\nonumber \\
\vec{\Upsilon}_{{ 12}} & = & \sigma_{0}\left(\tau_{1}\rho_{1}\mu_{1},\tau_{2}\rho_{1}\mu_{1},\tau_{0}\rho_{3}\mu_{0}\right)\nonumber \\
\vec{\Upsilon}_{{ 13-15}} & = & \left(\sigma_{0}\tau_{0}\rho_{3}\mu_{0},\sigma_{i}\tau_{1}\rho_{1}\mu_{2},\sigma_{i}\tau_{2}\rho_{1}\mu_{2}\right)\nonumber \\
\vec{\Upsilon}_{{ 16}} & = & \sigma_{0}\left(\tau_{1}\rho_{1}\mu_{3},\tau_{2}\rho_{1}\mu_{3},\tau_{0}\rho_{3}\mu_{0}\right)\nonumber \\
\vec{\Upsilon}_{{ 17}} & = & \left(\sigma_{1},\sigma_{2},\sigma_{3}\right)\tau_{3}\rho_{3}\mu_{0}\nonumber \\
\vec{\Upsilon}_{{ 18}} & = & \left(\sigma_{1}\tau_{3}\rho_{3}\mu_{0},\sigma_{2}\tau_{3}\rho_{3}\mu_{1},\sigma_{3}\tau_{3}\rho_{3}\mu_{1}\right)\nonumber \\
\vec{\Upsilon}_{{ 19}} & = & \left(\sigma_{1}\tau_{3}\rho_{3}\mu_{0},\sigma_{2}\tau_{1}\rho_{1}\mu_{2},\sigma_{3}\tau_{1}\rho_{1}\mu_{2}\right)\nonumber \\
\vec{\Upsilon}_{{ 20}} & = & \left(\sigma_{1}\tau_{3}\rho_{3}\mu_{0},\sigma_{2}\tau_{2}\rho_{1}\mu_{2},\sigma_{3}\tau_{2}\rho_{1}\mu_{2}\right)\nonumber \\
\vec{\Upsilon}_{{ 21}} & = & \left(\sigma_{1}\tau_{3}\rho_{3}\mu_{0},\sigma_{2}\tau_{0}\rho_{3}\mu_{2},\sigma_{3}\tau_{0}\rho_{3}\mu_{2}\right)\nonumber \\
\vec{\Upsilon}_{{ 22}} & = & \left(\sigma_{1}\tau_{3}\rho_{3}\mu_{0},\sigma_{2}\tau_{3}\rho_{3}\mu_{3},\sigma_{3}\tau_{3}\rho_{3}\mu_{3}\right)\nonumber \\
\vec{\Upsilon}_{{ 23}} & = & \left(\sigma_{1}\tau_{3}\rho_{3}\mu_{1},\sigma_{2}\tau_{3}\rho_{3}\mu_{0},\sigma_{3}\tau_{3}\rho_{3}\mu_{1}\right)\nonumber \\
\vec{\Upsilon}_{{ 24}} & = & \left(\sigma_{1}\tau_{1}\rho_{1}\mu_{2},\sigma_{2}\tau_{3}\rho_{3}\mu_{0},\sigma_{3}\tau_{1}\rho_{1}\mu_{2}\right)\nonumber \\
\vec{\Upsilon}_{{ 25}} & = & \left(\sigma_{1}\tau_{2}\rho_{1}\mu_{2},\sigma_{2}\tau_{3}\rho_{3}\mu_{0},\sigma_{3}\tau_{2}\rho_{1}\mu_{2}\right)\nonumber \\
\vec{\Upsilon}_{{ 26}} & = & \left(\sigma_{1}\tau_{0}\rho_{3}\mu_{2},\sigma_{2}\tau_{3}\rho_{3}\mu_{0},\sigma_{3}\tau_{0}\rho_{3}\mu_{2}\right)\nonumber \\
\vec{\Upsilon}_{{ 27}} & = & \left(\sigma_{1}\tau_{3}\rho_{3}\mu_{3},\sigma_{2}\tau_{3}\rho_{3}\mu_{0},\sigma_{3}\tau_{3}\rho_{3}\mu_{3}\right)\nonumber \\
\vec{\Upsilon}_{{ 28}} & = & \left(\sigma_{1}\tau_{3}\rho_{3}\mu_{1},\sigma_{2}\tau_{3}\rho_{3}\mu_{1},\sigma_{3}\tau_{3}\rho_{3}\mu_{0}\right)\nonumber \\
\vec{\Upsilon}_{{ 29}} & = & \left(\sigma_{1}\tau_{1}\rho_{1}\mu_{2},\sigma_{2}\tau_{1}\rho_{1}\mu_{2},\sigma_{3}\tau_{3}\rho_{3}\mu_{0}\right)\nonumber \\
\vec{\Upsilon}_{{ 30}} & = & \left(\sigma_{1}\tau_{2}\rho_{1}\mu_{2},\sigma_{2}\tau_{2}\rho_{1}\mu_{2},\sigma_{3}\tau_{3}\rho_{3}\mu_{0}\right)\nonumber \\
\vec{\Upsilon}_{{ 31}} & = & \left(\sigma_{1}\tau_{0}\rho_{3}\mu_{2},\sigma_{2}\tau_{0}\rho_{3}\mu_{2},\sigma_{3}\tau_{3}\rho_{3}\mu_{0}\right)\nonumber \\
\vec{\Upsilon}_{{ 32}} & = & \left(\sigma_{1}\tau_{3}\rho_{3}\mu_{3},\sigma_{2}\tau_{3}\rho_{3}\mu_{3},\sigma_{3}\tau_{3}\rho_{3}\mu_{0}\right)\nonumber \\
\vec{\Upsilon}_{{ 33}-35} & = & \left(\sigma_{0}\tau_{1}\rho_{1}\mu_{1},\sigma_{i}\tau_{1}\rho_{1}\mu_{2},\sigma_{i}\tau_{3}\rho_{3}\mu_{3}\right)\nonumber \\
\vec{\Upsilon}_{{ 36}} & = & \sigma_{0}\left(\tau_{1}\rho_{1}\mu_{1},\tau_{3}\rho_{3}\mu_{2},\tau_{1}\rho_{1}\mu_{3}\right)\nonumber \\
\vec{\Upsilon}_{{ 37-39}} & = & \left(\sigma_{0}\tau_{2}\rho_{1}\mu_{1},\sigma_{i}\tau_{2}\rho_{1}\mu_{2},\sigma_{i}\tau_{3}\rho_{3}\mu_{3}\right)\nonumber \\
\vec{\Upsilon}_{{ 40}} & = & \sigma_{0}\left(\tau_{2}\rho_{1}\mu_{1},\tau_{3}\rho_{3}\mu_{2},\tau_{2}\rho_{1}\mu_{3}\right)\nonumber 
\end{eqnarray}
\begin{eqnarray}
\vec{\Upsilon}_{{ 41-43}} & = & \left(\sigma_{0}\tau_{0}\rho_{3}\mu_{1},\sigma_{i}\tau_{0}\rho_{3}\mu_{2},\sigma_{i}\tau_{3}\rho_{3}\mu_{3}\right)\nonumber \\
\vec{\Upsilon}_{{ 44}} & = & \sigma_{0}\left(\tau_{0}\rho_{3}\mu_{1},\tau_{3}\rho_{3}\mu_{2},\tau_{0}\rho_{3}\mu_{3}\right)\nonumber \\
\vec{\Upsilon}_{{ 45-47}} & = & \left(\sigma_{0}\tau_{1}\rho_{1}\mu_{3},\sigma_{i}\tau_{1}\rho_{1}\mu_{2},\sigma_{i}\tau_{3}\rho_{3}\mu_{1}\right)\nonumber \\
\vec{\Upsilon}_{{ 48-50}} & = & \left(\sigma_{0}\tau_{2}\rho_{1}\mu_{3},\sigma_{i}\tau_{2}\rho_{1}\mu_{2},\sigma_{i}\tau_{3}\rho_{3}\mu_{1}\right)\nonumber \\
\vec{\Upsilon}_{{ 51-53}} & = & \left(\sigma_{0}\tau_{0}\rho_{3}\mu_{3},\sigma_{i}\tau_{0}\rho_{3}\mu_{2},\sigma_{i}\tau_{3}\rho_{3}\mu_{1}\right)\nonumber \\
\vec{\Upsilon}_{{ 54-56}} & = & \left(\sigma_{0}\tau_{3}\rho_{3}\mu_{2},\sigma_{i}\tau_{3}\rho_{3}\mu_{1},\sigma_{i}\tau_{3}\rho_{3}\mu_{3}\right)\label{eq:WZW_list-TBG}
\end{eqnarray}
Particularly interesting are those where at least some of the components
transform into each other under either SU$(2)_{{\rm spin}}$
or U$(1)$$_{{\rm valley}}$, i.e. $\vec{\Upsilon}_{{ 1}}$,
$\vec{\Upsilon}_{{ 12}}$, $\vec{\Upsilon}_{{ 16}}$
as well as $\vec{\Upsilon}_{{ 17}}$. The corresponding
superconducting states can be obtained with the help of Table~\ref{tab:TableTBG}.
Modulo a change of basis and taking into account that we only considered $\vec{\Upsilon}$ where all components transform the same under time reversal,  this  is closely related to Table III in Ref.~\onlinecite{Christos2020}. Differences in the two tables are due to the fact that  we do not group states into pairs w.r.t.~the mini-valley. This would remove $\vec{\Upsilon}_{{ 1}}$ and $\vec{\Upsilon}_{{16}}$ as options while adding a moir\'e density wave (second line in  Table III of   Ref.~\onlinecite{Christos2020}) as the third one. This third option can be directly obtained from Table~\ref{tab:TableTBG} as
\begin{equation}
    \vec{\Upsilon} = (\sigma_{0}\tau_{0}\rho_{3}\mu_{1},\sigma_{0}\tau_{3}\rho_{3}\mu_{2},\sigma_{0}\tau_{0}\rho_{3}\mu_{3})
\end{equation}
which maps to the associated line in Ref.~\onlinecite{Christos2020} upon basis transformation. The only compatible zero-momentum superconducting order parameter is the A$_1$ singlet.  



\end{document}